\documentstyle[12pt,epsfig]{article}
\def\bq{\begin{quote}} 
\def\eq{\end{quote}}

\def\bq{\begin{quote}} 

\def\bar{\overline}

\def\bar{\overline}

\catcode`\@=11
\def\@normalsize{\@setsize\normalsize{15pt}\xiipt\@xiipt
\abovedisplayskip 14pt plus3pt minus3pt%

\belowdisplayskip \abovedisplayskip
\abovedisplayshortskip  \z@ plus3pt%
\belowdisplayshortskip  7pt plus3.5pt minus0pt}
\def\small{\@setsize\small{13.6pt}\xipt\@xipt
\abovedisplayskip 13pt plus3pt minus3pt%
\belowdisplayskip \abovedisplayskip
\abovedisplayshortskip  \z@ plus3pt%
\belowdisplayshortskip  7pt plus3.5pt minus0pt
\def\@listi{\parsep 4.5pt plus 2pt minus 1pt
            \itemsep \parsep
            \topsep 9pt plus 3pt minus 3pt}}
 
\def\underline#1{\relax\ifmmode\@@underline#1\else
        $\@@underline{\hbox{#1}}$\relax\fi}
\@twosidetrue
\relax

\catcode`@=12

\catcode`\@=11

\def\section{\@startsection{section}{1}{\z@}{3.5ex plus 1ex minus
   .2ex}{2.3ex plus .2ex}{\large\bf}}

\def\ps@headings{\def\@oddfoot{}\def\@evenfoot{}
\def\@oddhead{\hbox{}\hfill
        \makebox[.5\textwidth]{\raggedright\ignorespaces --\thepage{}--
        \hfill }}
\def\@evenhead{\@oddhead}
\def\subsectionmark##1{\markboth{##1}{}}
}
 
\ps@headings
 
\catcode`\@=12
 
\relax

\def\figcap{\section*{Figure Captions\markboth
        {FIGURECAPTIONS}{FIGURECAPTIONS}}\list
        {Fig. \arabic{enumi}:\hfill}{\settowidth\labelwidth{Fig. 999:}
        \leftmargin\labelwidth
        \advance\leftmargin\labelsep\usecounter{enumi}}}
 \relax
\def\tablecap{\section*{Table Captions\markboth
        {TABLECAPTIONS}{TABLECAPTIONS}}\list
        {Table \arabic{enumi}:\hfill}{\settowidth\labelwidth{Table 999:}
        \leftmargin\labelwidth
        \advance\leftmargin\labelsep\usecounter{enumi}}}
 \relax
\def\reflist{\section*{References\markboth
        {REFLIST}{REFLIST}}\list
        {[\arabic{enumi}]\hfill}{\settowidth\labelwidth{[999]}
        \leftmargin\labelwidth
        \advance\leftmargin\labelsep\usecounter{enumi}}}
 \relax
 
\catcode`\@=11
 
\def\marginnote#1{}
\newcount\hour
\newcount\minute
\newtoks\amorpm
\hour=\time\divide\hour by60
\minute=\time{\multiply\hour by60 \global\advance\minute by-
\hour}
\edef\standardtime{{\ifnum\hour<12 \global\amorpm={am}%
    \else\global\amorpm={pm}\advance\hour by-12 \fi
    \ifnum\hour=0 \hour=12 \fi
    \number\hour:\ifnum\minute<100\fi\number\minute\the\amorpm}}
\edef\militarytime{\number\hour:\ifnum\minute<100\fi\number\minute}

\def\draftlabel#1{{\@bsphack\if@filesw {\let\thepage\relax
  \xdef\@gtempa{\write\@auxout{\string
    \newlabel{#1}{{\@currentlabel}{\thepage}}}}}\@gtempa
    \if@nobreak \ifvmode\nobreak\fi\fi\fi\@esphack}
     \gdef\@eqnlabel{#1}}
\def\@eqnlabel{}
\def\@vacuum{}
\def\draftmarginnote#1{\marginpar{\raggedright\scriptsize\tt#1}}
\def\draft{\oddsidemargin -.5truein
        \def\@oddfoot{\sl preliminary draft \hfil
        \rm\thepage\hfil\sl\today\quad\militarytime}
        \let\@evenfoot\@oddfoot \overfullrule 3pt
        \let\label=\draftlabel
        \let\marginnote=\draftmarginnote
 
\def\@eqnnum{(\theequation)\rlap{\kern\marginparsep\tt\@eqnlabel}%
\global\let\@eqnlabel\@vacuum}  }
\def\preprint{\twocolumn\sloppy\flushbottom\parindent 1em
        \leftmargini 2em\leftmarginv .5em\leftmarginvi .5em
        \oddsidemargin -.5in    \evensidemargin -.5in
        \columnsep 15mm \footheight 0pt
        \textwidth 250mmin      \topmargin  -.4in
        \headheight 12pt \topskip .4in
        \textheight 175mm
        \footskip 0pt
 
\def\@oddhead{\thepage\hfil\addtocounter{page}{1}\thepage}
        \let\@evenhead\@oddhead \def\@oddfoot{} \def\@evenfoot{}
}
\def\titlepage{\@restonecolfalse\if@twocolumn\@restonecoltrue\onecolumn
     \else \newpage \fi \thispagestyle{empty}\c@page\z@
        \def\thefootnote{\fnsymbol{footnote}} }
\def\endtitlepage{\if@restonecol\twocolumn \else  \fi
        \def\thefootnote{\arabic{footnote}}
        \setcounter{footnote}{0}}  
\catcode`@=12
\relax
 
\def\ps@headings{\def\@oddfoot{}\def\@evenfoot{}
\def\@oddhead{\hbox{}\hfill
        \makebox[.5\textwidth]{\raggedright\ignorespaces --\thepage{}--
        \hfill }}
\def\@evenhead{\@oddhead}
\def\subsectionmark##1{\markboth{##1}{}}
}
 
\ps@headings
 
\relax

\def\bo{{\raise.15ex\hbox{\large$\Box$}}}               
\def\face{{\raise.2ex\hbox{$\displaystyle \bigodot$}\mskip-2.2mu \llap {$\ddot
        \smile$}}}                                      

\def\leftrightarrowfill{$\mathsurround=0pt \mathord\leftarrow \mkern-6mu
        \cleaders\hbox{$\mkern-2mu \mathord- \mkern-2mu$}\hfill
        \mkern-6mu \mathord\rightarrow$}       
\def\dvec#1{\vbox{\ialign{##\crcr
        \leftrightarrowfill\crcr\noalign{\kern-1pt\nointerlineskip}
        $\hfil\displaystyle{#1}\hfil$\crcr}}}           

\def\beqx{\begin{displaymath}}
\def\eeqx{\end{displaymath}}

\newcommand{\newc}{\newcommand}
\newc{\ra}{\rightarrow}
\newc{\lra}{\leftrightarrow}
\newc{\beq}{\begin{equation}}
\newc{\eeq}{\end{equation}}
\newc{\bea}{\begin{eqnarray}}
\newc{\eea}{\end{eqnarray}}

\newc{\sm}{Standard Model}
\newc{\smd}{Standard Model}
\newc{\barr}{\begin{eqnarray}}
 \newc{\earr}{\end{eqnarray}}

\ps@headings
 
\relax
 
\def\firstpage#1#2#3#4#5#6{
\begin{document}


\begin{titlepage}
\nopagebreak
\title{\begin{flushright}
        \vspace*{-0.8in}
{\normalsize  hep-ph/9902364 \\  
ACT-1/99 \\
CERN-TH/99-16 \\
CTP-TAMU-04/99 \\
}
\end{flushright}
\vfill
{#3}}
\author{\large #4 \\[0.7cm] #5}
\maketitle
\vskip -7mm
\nopagebreak
\begin{abstract}
{\noindent #6}
\end{abstract}
\vfill
\begin{flushleft}
\rule{16.1cm}{0.2mm}\\[-3mm]

\end{flushleft}
\thispagestyle{empty}
\end{titlepage}}
 
\def\simlt{\stackrel{<}{{}_\sim}}
\def\simgt{\stackrel{>}{{}_\sim}}
\date{}
\firstpage{3118}{IC/95/34}
{\large \bf 
Leptogenesis in the Light of Super-Kamiokande Data and a
Realistic String Model
}
{John Ellis$^{\,a}$,  
S. Lola$^{\,a}$ and D.V. Nanopoulos$^{\,b,c,d}$}
{\normalsize\sl
$^a$Theory Division, CERN, CH 1211 Geneva 23, Switzerland\\[2.5mm]
\normalsize\sl
$^b$Center for Theoretical Physics, Department of Physics,\\[-1.0mm]
\normalsize\sl
  Texas A\&M
 University, College Station, TX 77843 4242,  USA\\[2.5mm]
\normalsize\sl
$^c$Astroparticle Physics Group, Houston Advanced Research Center (HARC),
\\[-1.0mm]
\normalsize\sl
The Mitchell Campus, Woodlands, TX 77381, USA\\[2.5mm]
\normalsize\sl
$^d$ Academy of Athens, Chair of Theoretical Physics, Division of Natural
Sciences,\\[-1.0mm]
\normalsize\sl
 28 Panepistimiou Ave., Athens GR-10679,  Greece.}
{We discuss leptogenesis
in the light of indications of neutrino masses and mixings from
Super-Kamiokande and other data on atmospheric neutrinos, as well as the
solar neutrino deficit.  Neutrino masses and mixings consistent with these
data may produce in a natural and generic way a lepton asymmetry that is
suffient to provide the observed baryon asymmetry, after processing via
non-perturbative electroweak effects.  We illustrate this discussion in
the framework of the string-derived flipped $SU(5)$ model, using particle
assignments and choices of vacuum parameters that are known to give
realistic masses to quarks and charged leptons.
We display one scenario for neutrino masses that also
accommodates leptogenesis.
}

\setcounter{page}{0}


\pagebreak

\section{Intoduction}
                                            
One of the basic questions in cosmology is the origin of the
observed baryon asymmetry of the Universe. This  could in principle
have arisen  either through non-perturbative
effects at the electroweak phase transition \cite{ref1},
or via lepton- and/or baryon-number-violating 
interactions at high temperatures.
Electroweak baryogenesis appears not to work in the
Standard Model, because, e.g., of the LEP lower limit
on the mass of the Higgs boson, but may be
possible in its supersymmetric extensions, though these
are also being constrained severely by LEP and other data.
Perturbative interactions that violate lepton and/or baryon
number arise naturally in grand unified extensions
of the Standard Model, and baryogenesis is actually one
of the main motivations for looking at these theories.
So far, no baryon-number-violating interactions have yet
been observed. However,
it has been pointed out that 
leptogenesis, e.g., via the out-of-equilibrium
decay of heavy Majorana neutrinos,
whose masses violate lepton number,
may lead to a net baryon asymmetry in the
universe \cite{fyan}, exploiting the fact that
lepton- and baryon-number-violating interactions are
expected to be in thermal equilibrium
at high temperatures. 
Within this approach \cite{rev}, it is found that the asymmetries of
baryon number $B$ and of $B-L$ are related by
\beq
     Y_B=\left({8N_f+4N_H\over22N_f+13N_H}\right)Y_{B-L}\,
\eeq
where $N_f$ is the number of quark-lepton families and $N_H$ the
number of Higgs doublets. This may easily yield the
required baryon asymmetry $ Y_B \equiv {n_B\over s} =
{n_B\over g_* n_\gamma} =
\sim 10^{-10}$.

There have recently been reports 
from  the Super-Kamiokande \cite{SKam} and other
\cite{Kam} collaborations, indicating the existence of
neutrino oscillations, for which the most natural mechanisms are
neutrino masses and mixings \cite{allref,ELLN2}.
In most models, the neutrino masses
are largely of Majorana type, which implies the existence of
interactions that violate lepton number. Thus, 
the physics beyond the Standard Model that
we are seeing for the first time may be just what we need to
generate the baryon asymmetry of the Universe.

One intriguing feature of 
the atmospheric-neutrino data is that they require a large neutrino
mixing angle~\cite{SKam,Kam}. Such mixing had  been shown,
before the SuperKamiokande data, to
arise naturally in a sub-class of GUT  models \cite{DG},
and more recently in certain models
with flavour symmetries \cite{OLDU1}.
Moreover, it is 
possible~\cite{ELLN2} to  accommodate the data 
in a natural and generic way within a flipped
$SU(5)\times U(1)$ model~\cite{aehn}
that is also consistent with the known 
hierarchies of charged-lepton and quark masses 
and mixings~\cite{ELLN}.
We showed in this analysis that flipped $SU(5)$ avoids the
tight relation
between $u$-quark and Dirac neutrino mass matrices
found in many GUTs, and includes $SU(3) \times SU(2) \times U(1)$-singlet
fields that are good candidates for $\nu_R$ fields.
With suitable choices of the parameters of the vacuum of the
string-derived flipped $SU(5)$ model,
we found  solutions to the
atmospheric neutrino deficit with 
a suitable hierarchy of neutrino masses. It was possible to
obtain either the small- or (perhaps preferably) the large-angle MSW
solution to the solar-neutrino problem, but not the `just-so'
vacuum-oscillation solution.

In this paper, we re-examine scenarios for leptogenesis~\cite{fyan,rev}
in the light of the new insights into neutrino masses
and mixings provided by the Super-Kamiokande and other data.
We find that leptogenesis can be successful if certain
supplementary constraints on the heavy Majorana neutrino mass
spectrum are obeyed. We illustrate these observations in the
framework of the flipped $SU(5)$ string model,
whose vacuum parameters are
constrained by both the quark and charged-lepton
mass hierarchies, as well as by the flat
directions of the theory. 
We find one scenario for neutrino
masses, within this general framework,
that is compatible with leptogenesis.


\section{Neutrino Masses, Reaction Rates and Boltzmann Equations}

We consider the generic likelihood that there is a hierarchy
of eigenvalues in the 
heavy Majorana neutrino mass matrix:
$M_{N_1} < M_{N_2},M_{N_3}$.
In such a case, the lightest right-handed neutrino
will usually still be in    equilibrium 
during the decays of the two heavier ones, therefore
washing out any lepton asymmetry generated by them.
For this reason, it is reasonable to assume that
any lepton asymmetry is generated only by the $CP$-violating decay of
    the lightest right-handed neutrino $N_1$. 

At tree level, the total decay width of $N_1$ 
\footnote{
Here, both the modes
$N_1 \rightarrow \phi^\dagger + \nu$
and $N_1 \rightarrow \phi + \bar{\nu}$,
where $\phi$ is the Higgs field, are included
.}
is given by
    \beq
\Gamma = \frac{(\lambda^\dagger \lambda)_{11}}{8 \pi} M_{N_1}
    \eeq
where $\lambda = m^D_\nu /v$, $v$ being the corresponding
light Higgs vacuum expectation value (vev). 
We do not assume that the Dirac neutrino couplings $\lambda$
are related directly to the $u$-quark Dirac couplings, as 
has often been assumed in previous works. As usual,
the leading contribution to
the $CP$-violating decay asymmetry, $\epsilon$, arises from the
interference between the tree-level decay amplitude and one loop amplitudes.
These include corrections of vertex type,
but may also involve self-energy corrections
$\tilde{\delta}$. The latter
may even be dominant if two of the heavy neutrinos are
almost degenerate \cite{FPS},
which can become the case in some of the examples that we study below.
In general, $\epsilon$ is given by
    \bea
     \epsilon_j & =  & 
{1\over( 8 \pi \lambda^{\dagger}\lambda)_{11}}
     \sum_j {\rm Im} 
\left [
(\lambda^\dagger \lambda)_{1j}^2
\right ] f \left (\frac{m^2_{N_j}}{m^2_{N_1}} \right )
\eea
where
\bea
 f(y) & = & \sqrt{y}\left[1-(1+y)\ln\left({1+y\over y}
     \right)\right]\;.
    \eea
We recall that, in order to calculate consistently
$CP$-violating asymmetries, lepton-number-violating scattering
processes must also be included. The complete cross sections for these
processes have been presented in \cite{luty}, where we refer the
reader for more details. 

On the other hand, $\tilde{\delta}$ roughly scales as
$1/\eta$, where $\eta = (M_{N_2} - M_{N_1})/ M_{N_1}$,
indicating that if the two masses are close in magnitude,
but not closer than the decay widths, 
a large enhancement of the 
lepton asymmetry may occur \cite{FPS}.
We note, however, that when
the two masses are exactly equal, the asymmetry vanishes 
as there is no mixing  between two identical
particles.

Let us define the variables
$Y=n/s$ and $x=M_{N_1}/T$, where
$n$ is the  number  of
neutrinos per co-moving volume element, whilst
$s$ is the entropy
density of the Universe. The latter is given by $s = g_* n_\gamma$,
where $g_*$ is the total number of spin degrees of freedom,
and $n_\gamma$ is the equilibrium photon density
of the Universe. We then have the following Boltzmann
equation for the time evolution of the neutrino number density:
\bea
{{\rm d}Y\over{\rm d}x}=- \frac{\Gamma(x) x}
{H(x=1)}  (Y-Y^{eq}),
\eea
where $H$ is the Hubble parameter and
\bea
Y^{eq}=n^{eq}/s=\left \{
\begin{array}{ll}
g_*^{-1} & x\ll 1 \\
g_*^{-1}\sqrt{\pi/2}x^{3/2}\exp(-x) & x\gg 1
\end{array}
\right. 
\eea
and we should impose the initial condition $Y(0)=
g_*^{-1}$.
The corresponding Boltzmann  equation for the lepton asymmetry $Y_L$ is
\bea
{{\rm d}Y_L\over {\rm d}x}= \epsilon \frac{\Gamma(x) x }
{H(x=1)} (Y-Y^{eq})-g_*Y^{eq}Y_L
\frac{\Gamma(x) x }
{2 H(x=1) }-{{2 Y_L\Gamma_s x}\over H(x=1)}
\eea
where  $\Gamma_s=n_\gamma<\sigma|v|>$, with the initial condition
$Y_L(0)=0$.
The lepton asymmetry at any time is given by
solving these coupled equations. Before doing so, however,
it is illuminating to see analytically what are
the direct constraints on the model parameters
that we can infer.

At the time of their decays, the neutrinos
have to be out of equilibrium, thus the decay rate $\Gamma$ has 
to be smaller than the Hubble parameter $H$ at temperatures $T\approx
M_{N_1}$.  $H$ is given by
\bea
H \approx 1.7 ~g_*^{1/2} ~\frac{T^2}{M_p}
\eea
where  in the Minimal Supersymmetric Standard model
$g_* \approx 228.75$, whilst $g_* = 106.75$ for the Standard
Model. This implies, as a first approximation, that
\bea
\frac{(\lambda^\dagger \lambda)_{11}}{14 \pi g_*^{1/2}} M_{p} < M_{N_1}
\label{outofeq}
\eea
However, a more accurate constraint is obtained by
looking directly at the solutions of the Boltzmann
equations for the system. Indeed, it turns out 
that even for Yukawa couplings larger than indicated in
(\ref{outofeq}),
the lepton-number-violating scatterings mediated
by right-handed neutrinos do not wash out completely 
the generated lepton asymmetry 
at low temperatures \cite{KolbTur}. Hence the
bound on the minimal value of $M_{N_1}$ in terms of the
Yukawa coupling is somewhat modified, 
as we discuss later in the analysis.

Demanding that the lepton asymmetry is generated before
the electroweak phase transition gives a constraint
on the lifetime of the right-handed neutrino, namely that it
has to be smaller than $10^{-12}\,$s. This implies that
\cite{plum}
    \beq
(m^{D\dagger}_\nu m^{D}_\nu)_{11}
>\left(20\,\mbox{eV}\right)^2
     \left({10^{10}\,\mbox{GeV}\over M_{N_1} }\right)\;
    \eeq
which is not a very severe constraint.

In a cosmological model with inflation, one has the additional
requirement that the decays of the right-handed neutrinos
should occur below the scale of inflation, which
is constrained by the magnitude of the density fluctuations
observed by COBE. This gives
\bea
M_{N_i} \leq m_{\eta} \leq 10^{13} ~\hbox{GeV}
\label{inflaton}
\eea
where $m_\eta$ is the inflaton mass, 
in generic inflationary models. However, this upper
limit may be increased by a couple of orders of magnitude
in models with preheating \cite{preheating}.

Finally, one has also to take into account 
the likelihood that the lepton asymmetry
produced in this framework is diluted by subsequent
entropy production~\footnote{For example, this may
take place during the breaking of $SU(5) \times U(1)$,
in the model discussed below.}. This is discussed later in
the paper.

Let us now incorporate the constraints from neutrino masses and mixings.
The Super-Kamiokande as well as the solar neutrino data,
which require small mass differences,
can be explained by two possible neutrino hierarchies:

(a) Textures with almost 
degenerate neutrino mass eigenstates, of the
order of ${\cal O} ({\rm eV})$.  In this case neutrinos may
also provide a component of hot dark matter.

(b) Textures with 
large hierarchies of neutrino masses:
$m_{\nu_3} \gg m_{\nu_2}, m_{\nu_1}$,
leaving open the
possibility of a second hierarchy
$m_{\nu_2} \gg m_{\nu_1}$. 
Then, the atmospheric neutrino data
requires
$m_{\nu_3} \approx (10^{-1} \; {\rm to} \;
10^{-1.5})$ eV
and $m_{\nu_2} \approx (10^{-2} \; {\rm to} \;
10^{-3})$ eV.

This data, clearly constraints the possible 
mass scales of the problem.
The mass of the heavier neutrino
is given by
\begin{eqnarray}
m_{\nu_3} = 
\frac{ (m_\nu^D)_{33}^2}{M_{N_3}}
\end{eqnarray}
For a scale ${\cal O} (200 ~ {\rm GeV})$ for the
Dirac mass, one has the following:
Solutions of the type (a), that is
light neutrinos of almost equal mass,
require 
\bea
M_{N_3}  \approx {\cal O}({\rm a ~few ~times ~10^{13}} ~ {\rm GeV})
\label{con1}
\eea
However, given that the Dirac neutrino couplings  are
expected in many unified or partially
unified theories to have large
hierarchies (similar to those of quarks), we conclude that
in order to obtain three almost degenerate neutrinos, a
large hierarchy
in the heavy Majorana sector would also be required.
(We emphasize that this is true in the case that no reverse Dirac
hierarchies are generated. An exception to this will
be the example we give subsequently, where there are large entries in
the 1-2 sector of the Dirac neutrino texture).
In the simple case of standard neutrino
Dirac hierarchies, the scale $M_{N_1}$ will be expected to be
significantly lower than the upper bound on the
inflaton mass.

On the other hand, solutions of the type (b),
with large light neutrino hierarchies
require 
\bea
M_{N_3} \approx O({\rm a ~few ~times ~10^{14}-10^{15} ~{\rm GeV}})
\label{con2}
\eea
Then, the inflaton mass condition demands
heavy Majorana hierarchies of the type
\begin{eqnarray}
\frac{M_{N_1}}{M_{N_3}}  \leq {\cal O} 
\left (
\frac{1}{100} \right ) <  \frac{M_{N_2}}{M_{N_3}}
\end{eqnarray}

The suppression of
$m_{\nu_2}$ with respect to $m_{\nu_3}$ 
(which is roughly 1/10)
can again be obtained either from the Yukawa couplings,
or from the heavy Majorana mass hierarchies:
For $M_{N_2} \approx M_{N_3}$ the relevant squared
Yukawa couplings should have a ratio 1:10.
However, for $M_{N_2} < M_{N_3}$ the ratio
of the relevant squared Yukawa couplings
has to be larger. The same is true for the 
relative suppression of
 $m_{\nu_1}$ with respect to $m_{\nu_2}$.
Here, however, the data offers no information on
how large  $m_{\nu_1}/m_{\nu_2}$ can be
(although the most natural expectation would be that
there is a second large hierarchy). 
However, in case (b), $M_{N_1}$
can be close to the inflaton mass,
unlike what happens in case (a).

It is interesting to note that leptogenesis
does not allow for reverse hierarchies in the
heavy Majorana mass sector, consistent with
the neutrino data, even in the case 
that the Yukawa couplings would be close in
magnitude. Indeed, the scales given by
eqs. (\ref{con1}) and (\ref{con2}), in the
case of reverse hierarchies, can never
be consistent with the bound  on the lightest
heavy neutrino mass scale from the inflaton mass.
At this stage it is difficult to 
obtain any additional 
information, without entering in more detail in the 
structure (and in particular the mixings) 
of the various mass matrices. 
This will be done in the next section, in
the framework of a realistic example.
However, from the above discussion it is clear that
leptogenesis provides an additional
probe to neutrino mass hierarchies.

At this stage, it is instructive to
illustrate how the lepton asymmetry evolves in the presence of 
rescattering in such a scenario.
In principle, one expects the following \cite{evol}:
for small Dirac neutrino couplings and a large
scale $M_{N_1}$, the scattering  that
tends to deplete the lepton asymmetry is suppressed.
In this case, the lepton asymmetry grows to a constant
asymptotic value, $Y_{asym}^1$. On the other hand, for larger
Yukawa couplings and smaller $M_{N_1}$, the scattering
processes start becoming relevant. Consequently,
the lepton asymmetry exhibits an increase to a peak, followed by
a subsequent decrease to an asymptotic value
$Y_{asym}^2 < Y_{asym}^1$. From the previous discussion, 
we see that the rough estimate of
the out-of-equilibrium condition (\ref{outofeq}), for
$M_{N_1} < 10^{13} ~{\rm GeV}$ (\ref{inflaton}),
corresponds to a bound
\beq
(\lambda^\dagger \lambda)_{11} < 6 \cdot 10^{-4}
\label{boundlambda}
\eeq
For any $(\lambda^\dagger \lambda)_{11}$ below this
value, the lepton asymmetry grows to a constant value.
However, it turns out that we may allow higher values,
and still get a large enough lepton asymmetry,
as is illustrated in Figure 1.

\begin{figure}[tbp]
\vspace*{-5.1 cm}
\centerline{\epsfig{figure=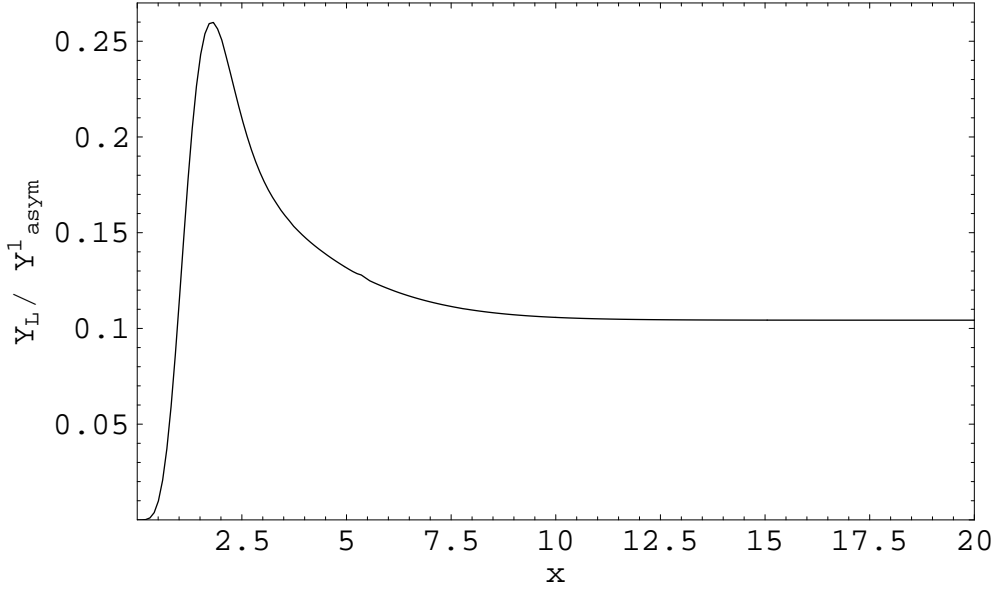,width=1.2\textwidth,clip=}}
\vspace*{-13.0 cm}
\caption{{\it Evolution of lepton asymmetry for 
$M_{N_1}\approx 10^{13}$~{\rm GeV} and $(\lambda^\dagger \lambda)_{11} =
1.4 \cdot 10^{-3}$.}}
\end{figure}

What is the final baryon asymmetry that is generated?
This is given by 
\bea
Y_B \equiv \frac{n_B}{s} \sim  \frac{n_L}{s} \sim 
\left ( \frac{m_{\eta}}{M_p} \right )^{1/2} 
\frac{Y_{asym}^2}{Y_{asym}^1} ~\frac{(\epsilon +\tilde{\delta})}{g_*}
~\frac{1}{\Delta} 
\eea
where $\Delta$ is a dilution factor, due to 
entropy that is produced during the breaking 
of $SU(5)\times U(1)$ \cite{cehno1} when a singlet
field $\Phi$ (flaton) gets a vev. This is given by \cite{JD}
\begin{equation}
\Delta = \frac {s(R_{d\Phi})}{s(R_{d\eta})}
(\frac {R_{d\Phi}}{R_{d\eta}})^3  \sim
\frac {V^3 m_\eta^{3/2}}{{\alpha_\Phi}^{1/2}
\ m_{SUSY}^{3/2} M_P^3} 
\label{entropyfactor}
\end{equation}
In the above, the $\Phi$ decay rate is given by
$\Gamma_\Phi = \alpha_\Phi 
\frac { m^3_{SUSY}}{V^2}$, $V$ is the scale where
the vev of the Higgs ${\tt 10}$ and ${\tt \bar{10}}$ 
break the flipped-SU(5) group,
and we take the supersymmetry breaking scale 
to be $m_{SUSY} \sim 10^{-16}M_P$.
We will discuss below which is the
magnitude of the dilution factor that one may accommodate
in a successful scheme of leptogenesis.

\section{Specialization to a Realistic Flipped $SU(5)$ String Model}

As an exemplar of the above combined analysis of
leptogenesis and phenomenological constraints,
we consider the `realistic'
flipped $SU(5)$ model derived from string, working with the mass matrices
discussed in~\cite{ELLN,ELLN2}.
Relevant aspects of this model are reviewed
in the Appendix: it contains 
many singlet fields, and the mass matrices depend on the subset of these
that get non-zero vev's, i.e., on the 
choice of flat direction in the effective potential.
The questions we study in this section are: is the choice
of vev's made previously consistent with the cosmological
constraints discussed in the previous section? and, if so:
does cosmology further constrain the model parameters in an
interesting way?

Within this model, we found that the charged-lepton mixing matrix 
is given by
\beq
V^m_{\ell_L} = \left (
\begin{array}{ccc}
1-\frac{1}{2} (\Delta_2 \Delta_5)^2 & \Delta_2 \Delta_5 & 0 \\
-\Delta_2 \Delta_5  &  1-\frac{1}{2} (\Delta_2 \Delta_5)^2 & 0 \\
0 & 0 & 1
\end{array}
\right )
\label{veenum}
\eeq
where $\Delta_2 \Delta_5$ is a combination of 
hidden-sector fields that transform as
sextets under $SO(6)$ (see the Appendix for the relevant field
definitions).
In the same framework, $m_{\nu}^D$ was found to take  the form
\begin{equation} 
m_{\nu}^D = 
\left (
\begin{array}{ccc}
\Delta_2 \Delta_5 \bar{\phi}_4 & 1 & 0 \\
\bar{\phi}_4 & \Delta_2 \Delta_5 & 0 \\
0 & 0 & F_1
\end{array}
\right) \equiv
\left (
\begin{array}{ccc}
x f & 1 & 0 \\
f & x & 0 \\
0 & 0 & y
\end{array}
\right) 
\label{Dira}
\end{equation}
where $F_1$ and $\bar{\phi}_4$ are fields also defined
in the Appendix, whose vev's are going to determine the
magnitudes of the various entries.
The form of the heavy Majorana mass matrix, $M_{\nu_R}$,
is found to be \footnote{
Here, we use the notation of \cite{ELLN2}.}
\begin{equation} 
M_{\nu_R} = 
\left (
\begin{array}{ccc}
\bar{F}_5 \bar{F}_5  \bar{\phi}_4 \phi_3 &
\bar{F}_5 \bar{F}_5  \Delta_2 \Delta_5 \phi_3 & 0 \\
\bar{F}_5 \bar{F}_5  \Delta_2 \Delta_5 \phi_3 &
0 & \bar{F}_5 \bar{\Phi}_{31}\Phi_{31} \bar{\phi}_4 \phi_2 \\
0 & \bar{F}_5 \bar{\Phi}_{31}\Phi_{31} \bar{\phi}_4 \phi_2 
& \Delta_2 \Delta_5 \bar{\Phi}_{23} T_2 T_5
\end{array}
\right)
\equiv
\left (
\begin{array}{ccc}
M & M' & 0 \\
M' & 0 & M_{4\phi} \\
0 & M_{4\phi} & M_{\phi \phi}
\end{array}
\right )
\label{hema}
\end{equation}
Using the above formulae, we were able to
calculate the light-neutrino mass matrix,
which is given by the standard see-saw formula:
\begin{equation}
m_{eff}=m^D_{\nu}\cdot (M_{\nu_R})^{-1}\cdot (m^{D }_{\nu})^T
\label{eq:meff}
\end{equation}
in terms of the Dirac neutrino mass matrix $m_{\nu}^D$ 
and the heavy Majorana neutrino mass matrix $M_{\nu_R}$
introduced above.
We note, moreover, that the mixing in the leptonic sector is
given by
\beq
V_{\nu} = V^{m \dagger}_{\nu} V^m_{\ell_L}
\label{bothmix}
\eeq
where the symbols $V^m_{\nu}, V_{\ell_L}$ denote the 
flavour-rotation
matrices for neutrinos and left-handed charged leptons,
respectively, required to diagonalize their mass matrices.

We see that, in this model, potentially large off-diagonal entries 
in the heavy Majorana mass matrix 
may yield large neutrino mixing.  Moreover, 
the neutrino Dirac matrix, which is {\it not} equivalent to $m_u$
in this model,
also provides a potential source of large $\nu_{\mu} -
\nu_{e}$ mixing.

Clearly, the forms of the mass matrices depend
on the various field vev's. For these, we have already used
information from the analysis of the flat directions and the
fermion masses. We recall that
our analysis of quark masses pointed towards
$\Delta_2 \Delta_5 = {\cal O} (1)$,
as well as a suppressed value of $\bar{\phi}_4 \ll 1$.
Moreover, from the analysis of flat directions~\cite{ELLN},
we concluded that $\bar{\Phi}_{31} \bar{\Phi}_{23} = {\cal O} (1)$ 
is large. In addition,
the flatness conditions~\cite{ELLN} 
relate $\bar{\Phi}_{31},\Phi_{31}$ and
$\phi_2$, and can be satisfied even if all
the vev's are large, as long as
$\bar{\Phi}_{31} \Phi_{31}$ and
$\bar{\Phi}_{23} \Phi_{23}$ are
not very close to unity.
As for the
decuplets that break the gauge group down to
the Standard Model, we know that the vev's 
should be 
$\approx M_{GUT}/M_{s}$. In weakly-coupled string constructions,
this ratio is $\approx 0.01$. However, the strong-coupling limit of
$M$ theory offers the possibility that the
GUT and the string scales can coincide, in which case the
vev's could be of order unity. 

On the other hand, flatness conditions and quark
masses do not give any information on the vev of the product
$T_2 T_5$. Even this combination, however, is constrained from the
requirements for the light neutrino masses \cite{ELLN2}.
Finally, the field $\phi_3$ is the one for which we seem
to know least and we will discuss in a subsequent
section how its value may affect leptogenesis.

\subsection{First Class of Solutions}

In \cite{ELLN2}, where we classified the
flipped $SU(5)$ solutions to the super-Kamiokande data,
we first considered the following simplified form
for the heavy Majorana mass matrix:
\bea
M_{\nu_R} \propto
\left (
\begin{array}{ccc}
M & 0 & 0 \\
0 & 0 & M_{4\phi} \\
0 & M_{4\phi} & M_{\phi\phi}
\end{array}
\right )
\equiv 
\left (
\begin{array}{ccc}
M & 0 & 0 \\
0 & 0 & f y \\
0 & f y & t x
\end{array}
\right )
\label{simplems}
\eea
where our approximation was to neglect $M'$ - but not to make
any other {\it a priori} assumption about the relative magnitudes
of entries in $M_{\nu_R}$. 
This approximation can be motivated if
$\phi_3$ is negligible~\cite{ELLN}, and $M$ is eventually
generated by some other effect.
Then, we showed that the magnitude of
$M$ is not essential for the
calculations of the light neutrino data.
For leptogenesis however the situation will be much different,
the reason being that $M$ is 
{\em directly associated} with the lighter eigenvalue of
the Heavy Majorana mass sector.
Since $M$ essentially decouples from the rest of the
entries, this is the easiest
example one can calculate, since we can essentially
read off the masses and Yukawa couplings that
we need without explicitly calculating any
mixing matrices. This will not be the case in the
next section, where we will need to transform the
Dirac neutrino mass matrices in the basis where
the heavy Majorana one is diagonal.

We also showed that
consistency within this framework, required ${\bar F}_5$ to be quite large,
as could occur in the strong-coupling limit of $M$ theory. Finally,
in this scheme the combination $T_2 T_5 $
was also fixed to be $ T_2 T_5 \sim {\bar \phi}_4$
\footnote{The actual value of
$M$ is irrelevant for $m_{eff}$, provided
$M$ is larger than $\approx 
(\bar{\phi}_4^4 \Delta_2 \Delta_5  F_1^2)/(T_2 T_5)$
in normalised quantities.}.

Let us now go to the Dirac mass matrix. We saw
that for the calculation of the lepton asymmetry, 
we need to know the combination $(\lambda^{\dagger} \lambda)_{11}$.
This can be read by
\bea
(m_{\nu}^{D\dagger}  m_{\nu}^D) \propto 
\left (
\begin{array}{ccc}
f^2 (1+x^2) & 2 f x & 0 \\
2 f x & 1+x^2 & 0 \\
0 & 0 & y^2
\end{array}
\right)
\label{DiraSq}
\eea
This indicates that 
$(\lambda^{\dagger}  \lambda)_{11}$ is suppressed
as compared to 
$(\lambda^{\dagger}  \lambda)_{22}$ and
$(\lambda^{\dagger}  \lambda)_{33}$, which for strong unification
are of the same order of magnitude.
The eigenvalues of the heavy Majorana mass matrix are:
\bea
M_{N_1} & = & M \nonumber \\
M_{N_2} & = & \frac{1}{2} (tx -\sqrt{t^2 x^2 + 4 f^2 y^2 }) \nonumber \\
M_{N_3} & = & \frac{1}{2} (tx + \sqrt{t^2 x^2 + 4 f^2 y^2 }) 
\eea
thus allowing for the possibility of a hierarchy
between 
$M_{N_2}$ and $M_{N_3}$.

We can now use the above hierarchies, in order to estimate 
what would be the natural magnitude for $\epsilon$.
The above  Yukawa couplings and masses indicate that
the dominant contribution arises from second-generation particles,
in the decays of the heavy neutrinos of the first-generation.
Then,
\begin{equation}
\epsilon_{12} \approx {1 \over 8\pi} 
~\frac{(\lambda^{\dagger} \lambda)_{12}^2}
{(\lambda^{\dagger} \lambda)_{11}}
 ~ f \left (
{M_2^2 \over M_1^2}\right ) ~\delta
\end{equation}
where $\delta$ is the CP-violating phase factor.
Depending how close 
$M_2$ is to  $M_1$, $\epsilon$ may be as large as
$ 10^{-2} \delta$~\footnote{ In the extreme case that 
$M_2$ is very close to $M_1$,
one would in principle have to consider the
evolution of the coupled equations
for the two neutrinos.  However, in our solutions,  the second 
neutrino has a large coupling that brings it in equilibrium.
Consequently, it is only one neutrino  that finally contributes
to the lepton asymmetry of the universe, even in this 
case .}. On the other hand, $\tilde{\delta}$ may be significantly
enhanced for $M_1 \approx M_2$, although for large mass differences
it is of the same order as $\epsilon$.

Finally, we need to calculate the ratio
$(Y_{asymm}^2/Y_{asymm}^1)$ from the Boltzmann 
equations and for $(\lambda^\dagger \lambda)_{11}
\approx f^2 \approx 0.0016$
(where we stress again that we neglect coefficients
of order unity, which are currently not
predicted by the theory).
It turns out that, for $M_{N_1} = 10^{13} ~{\rm GeV}$,
\bea
\frac{Y_{asymm}^2}{Y_{asymm}^1} =  9 \cdot 10^{-2}
\eea 
whilst, for $M_{N_1} = 10^{11} ~{\rm GeV}$,
\bea
\frac{Y_{asymm}^2}{Y_{asymm}^1} = 4 \cdot 10^{-4}
\eea 
We see therefore that by lowering $M_{N_1}$
while keeping the Yukawa coupling fixed, we 
significantly lower the produced lepton asymmetry.
However, remember that a higher 
value of $M_{N_1}$ also requires a higher inflaton
mass and then the dilution factor $\Delta$ becomes
larger.

We see, then, that 
(even in the case that $M_{N_1}$ and $M_{N_2}$,
while being close in magnitude, 
do not fulfil the resonant condition that increases
the generated asymmetry)
$M_{N_1} = 10^{13}$~GeV yields a ratio
\bea
Y_B \equiv  \frac{n_B}{s} \approx
~\frac{ 3 \cdot 10^{-8}}{\Delta} 
\eea
which can be in the acceptable range
$Y_B^{observed}  \approx  10^{-10}$ for
$\Delta \leq  300 $. This is difficult to reconcile with
(\ref{entropyfactor}),
but might be consistent with suitably 
large $\alpha_{\Phi}$ and $m_{SUSY}$.~\footnote{The
general issue of flaton decay may need to be reviewed in the new
$M$-theory context of strongly-coupled string.}

Moreover, note that the various entries in the
mass matrices, are only known {\em up to order unity
coefficients}, while  a small change in a Yukawa 
coupling can have a large effect on the ratio
$(Y_{asymm}^2/Y_{asymm}^1)$. Indeed,
for $M_{N_1} = 10^{13}$ GeV, one has the following:
for $(\lambda^\dagger \lambda)_{11} = 0.001$,
\bea
\frac{Y_{asymm}^2}{Y_{asymm}^1} = 0.16
\eea
whilst, for $(\lambda^\dagger \lambda)_{11} = 0.0004$,
\bea
\frac{Y_{asymm}^2}{Y_{asymm}^1} =  0.45
\eea
Finally, we recall that in the presence
of preheating, one may raise the limit on the inflaton
mass, and hence the value of
$Y_{asymm}^2 /Y_{asymm}^1$. For example, if 
$M_{N_1} = 10^{14}$ GeV and $(\lambda^\dagger \lambda)_{11} = 0.001$,
we find
\bea
\frac{Y_{asymm}^2}{Y_{asymm}^1} =  0.86
\eea
However, since 
\bea
Y_B ~  \propto ~\frac{1}{\Delta} \left (
\frac{m_{\eta}}{M_p}
\right )^{1/2}
\sim \frac{1}{m_{\eta}} 
\eea
by raising the limit on the inflaton mass and thus
the possible mass for the lighter neutrino,
we end up with a larger suppression due to entropy
production.

\subsection{Second Class of Solutions}

We will now investigate whether our parametrization of the
flipped $SU(5)$ model matches the cosmological requirements,
for the second class of solutions that we found in
\cite{ELLN2}. These occur in 
the case that 
the field $\phi_3$ develops a large vev.
Previously, we had  assumed for
simplicity that $\phi_3\approx 1$.
This is actually the most natural range, 
given that large 
$\phi_3$ allows for a suppression of
$m_{\nu_{e}}$ and $m_{\nu_{\mu}}$
as compared to $m_{\nu_{\tau}}$.
Then,  we 
can write $M_{\nu_R}$ in the form
\begin{equation} 
M_{\nu_R} \propto 
\left (
\begin{array}{ccc}
f y^2  & 2 x y^2  & 0 \\
2 xy^2  & 0 & f  y \\
0 & f y & t x
\end{array}
\right) \label{maj}
\end{equation}
where
$\Delta_2 \Delta_5 \equiv x, T_2 T_5 \equiv t$,
$\bar{\phi}_4 \equiv f$
and $\bar{F}_5 \equiv y$.
In the above, the factor of $2$ has been included
in order to avoid {\em sub-determinant cancellations},
which are not expected to arise once order unity
coefficients are properly taken into account.

Let us then write down $m_{eff}$ in the 
flipped-$SU(5)$ field basis.
This is given by
\bea
m_{eff} \propto \left ( 
\begin{array}{ccc}
-f^4 x^2 + tfx & -f^4 x - tf x^2 & -f^2 y^2  \\
 -f^4 x - tf x^2 &  -f^4 - 3 f t x^3 & f^2 x y^2 \\
-f^2 y^2 & f^2 x y^2 & -4 x^2 y^4 
\end{array}
\right )
\eea
As we see from this matrix 
(and have stressed in our previous analysis),
the neutrino data 
solutions with large light neutrino
hierarchies
require  $\bar{\phi}_4 \approx \bar{F}_5,F_1$,
as in weak-coupling unification schemes \cite{ELLN2}.
On the other hand,  $T_2 T_5$ is not
fixed to a specific value, however it 
has to be smaller than $(\bar{\phi}_4)^3$, so that
the entries in the (1,2) sector of $m_{eff}$
remain small. Here we should stress that 
{\em ${\cal O} {\rm (1)}$ coefficients  may not be fixed}
by the model  and therefore we are only concerned 
with the order of magnitude of the various entries,
as it is specified by the operators.

Finally, the Dirac mass matrix, is similar to 
the one calculated before, with the difference that
$y$ is now much smaller.
In this example, the light Majorana mass matrix does
not decouple and therefore in order to work with
the $ M_{N_i}$ mass eigenstates, we need to 
diagonalise $M_{\nu_R}$ and also
transform $m_{\nu}^D$  to the  basis where
$M_{\nu_R}$ is diagonal.
Indeed, let
\bea
M_{\nu_R}^{diag} = V^T \cdot M_{\nu_R} \cdot V
\eea
Then the Yukawa couplings have to be calculated from the
matrix
\bea
\tilde{m}_{\nu}^D ~=~ {m}_{\nu}^D \cdot V
\eea

Since in this class of solutions we require
$x \approx 1, y \approx f $ and $ t \leq f^3$, we can express the
solutions only in terms of the parameter $f$.
Let us first calculate the eigenvalues of the 
heavy Majorana mass matrix. For the particular
choice of coefficients that appear in 
eq.(\ref{maj}), these 
scale as $ f^3/5 : \sqrt{5} f^2:-\sqrt{5} f^2$.
Note here that the coefficients are not so relevant,
since we do
not have any information about $O(1)$ factors from the
model; given that a small difference in a mass entry
may lead to a significantly larger factor in the eigenvalues,
we see that there is some room for arbitrariness.
What is unambiguous however, is that in this class of solutions,
the lightest eigenvalue tends to be suppressed as to the
heavier ones, by a factor of $f \approx 0.04$,
while  the two heavier eigenvalues are
of almost equal magnitude.

The mixing matrix, again for the
coefficient choice of eq.(\ref{maj}) and
keeping only the dominant contributions is
given by
\bea
V = 
\left (
\begin{array}{ccc}
0.63 & 0.63 & -0.45 \\
0.70 & -0.70 & 0.18 ~f \\
0.32 & 0.32 & 0.89 
\end{array}
\right )
\eea
where we see that in this example,
almost all the dominant entries of the mixing matrix are of order unity.
This was to be expected, since the dominant entries in 
$M_{\nu_R}$ are the off-diagonal ones.
Of course, the exact value of the mixing depends
on unknown coefficients, however 
since it is nearly maximal, suppression factors
of the order of $\sqrt{2}/2 \approx 0.70$ arise in any case.

The above discussion implies that 
the large off-diagonal factors in $m_\nu^D$ will start getting
communicated in the diagonal ones.
Indeed (for $x \approx 1, y \approx f $),
\bea
\tilde{m}_{\nu}^D ~\approx~ 
\left (
\begin{array}{rrr}
0.70 & - 0.70 & -0.27 f \\
0.70 & -0.70 & -0.27 f \\
0.32 f & 0.32 f & 0.9 f 
\end{array}
\right )
\eea
and therefore
\bea
\tilde{m}_\nu^{D \dagger} \tilde{m}_\nu^{D }
 = 
\left (
\begin{array}{ccc}
1 & -1 & -0.4 f \\
-1 & 1 & 0.4 f \\
-0.4 f & 0.4 f & 0.9 f^2 
\end{array}
\right )
\eea
thus indicating a significant increase in 
$(\lambda^{\dagger} \lambda)_{11}$ , 
a small decrease in $(\lambda^{\dagger} \lambda)_{22}$ 
and a larger decrease in
$(\lambda^{\dagger} \lambda)_{33}$, which are
in the wrong direction for leptogenesis.
This combined with the suppression of 
the second lightest eigenvalue with respect to
the lighter one,  which
reduces the value of $\epsilon$ and
thus of $Y_B$ by a factor of
$\approx 0.04$,
seems to make this case not viable.

Suppose now that we leave the field $\phi_3$ 
as a free parameter. Then, the heavy Majorana mass matrix 
becomes of the form
\begin{equation} 
M_{\nu_R} \propto 
\left (
\begin{array}{ccc}
f y^2 \phi_3 & 2 x y^2 \phi_3 & 0 \\
2 xy^2 \phi_3  & 0 & f  y \\
0 & f y & t x
\end{array}
\right) \label{maj2}
\end{equation}
and the light effective one
\bea
m_{eff} \propto \left ( 
\begin{array}{ccc}
-f^4 x^2 + tfx\phi_3 & -f^4 x - tf x^2 \phi_3 & -f^2 y^2 \phi_3 \\
 -f^4 x - tf x^2 \phi_3 &  -f^4 - 3 f t x^3 \phi_3 & f^2 xy^2 \phi_3 \\
-f^2 y^2 \phi_3 & f^2 x y^2 \phi_3 & -4 x^2 y^4 \phi_3^2
\end{array}
\right )
\eea

Then, we see that viable neutrino hierarchies are also obtained
for strong unification $(y \approx 1)$ and 
$\phi_3 \approx f^2$. The eigenvalues of
$m_{\nu_R}$ scale as $f^2: 1:-1$, while
the mixing matrix is now
\bea
V = 
\left (
\begin{array}{ccc}
0 & 0 & -1 \\
0.7 & -0.7 & 0 \\
0.7 & 0.7 & 2 f
\end{array}
\right )
\eea
Then (for $x,y \approx 1$)
\bea
\tilde{m}_{\nu}^D ~\approx~ 
\left (
\begin{array}{ccc}
0.7 & -0.7 & -f  \\
0.7  & -0.7  & -f \\
0.7 & 0.7 & 2 f
\end{array}
\right )
\eea
and
\bea
\tilde{m}_\nu^{D \dagger} \tilde{m}_\nu^{D }
 = 
\left (
\begin{array}{ccc}
1.5 & -0.5 & 0 \\
-0.5 & 1.5 & 3 f \\
0 & 3f & 6 f^2 
\end{array}
\right )
\eea
indicating that in this limit of field vevs as well,
the coupling $(\lambda^\dagger \lambda)_{11}$ is
large, thus not allowing the fulfilment of the
out-of-equilibrium condition.

\section{Conclusions}

In this paper, we have revisited leptogenesis in the light of the
indications from Super-Kamiokande~\cite{SKam} and other data~\cite{Kam} 
for non-zero neutrino
masses. These data provide significant additional constraints on scenarios
for leptogenesis~\cite{fyan,rev}, in particular on the possible magnitudes
of heavy
Majorana masses and on the possible patterns of mixing. We have shown
that a plausible framework for leptogenesis is compatible with these
new experimental constraints.

Our discussion has been illustrated by examples derived from a
flipped $SU(5)$ string model for quark and charged-lepton
masses~\cite{ELLN}, which we extended recently~\cite{ELLN2} to models of
neutrino masses
that were compatible with the Super-Kamiokande and other data.
We have shown that one of the neutrino-mass models proposed
previously leads to an acceptable realization of leptogenesis,
whilst the other has problems. This analysis exemplifies the
power of leptogenesis to refine further the selection of
realistic models. We consider it a non-trivial success of
flipped $SU(5)$ that it may survive this new set of constraints.

The more general message for model-builders that we extract from
this analysis is that one must be wary about the couplings
between the first generation and the other two. If there is a large
off-diagonal Dirac-type Yukawa coupling, as may arise in flipped
$SU(5)$, the mixing between the first-generation and other heavy
Majorana masses is constrained. The emerging pattern of light
neutrino masses suggests that the first and other two generations may have
substantial mixing, which could arise {\it a priori} from either the Dirac
and/or the heavy Majorana sectors. Our analysis suggests that
leptogenesis may be unhealthy if one combines the two sources
of mixing. The essential reason for this is the
out-of-equilibrium condition on the neutrino couplings.

\vspace*{0.15 cm}

{\bf Acknowledgements}

The work of D.V.N. has been supported in part by the U.S. Department of
Energy under grant DE-FG03-95-ER-40917.

\normalsize

\vspace*{0.15 cm}

\begin{center}
{\bf Appendix}
\end{center}
In this appendix we tabulate for completeness
the field assignment of the `realistic'
flipped $SU(5)$ string model~\cite{aehn},
as well as the basic conditions used in~\cite{ELLN} to obtain consistent
flatness conditions and acceptable Higgs masses.

\begin{table}[h]
\begin{center}
\begin{tabular}{|l||l||l|}
\hline
$F_1(10,\frac{1}{2},-\frac{1}{2},0,0,0)$ &
$\bar{f}_1(\bar{5},-\frac{3}{2},-\frac{1}{2},0,0,0)$ &
$\ell_1^c(1,\frac{5}{2},-\frac{1}{2},0,0,0)$ \\
 
$F_2(10,\frac{1}{2},0,-\frac{1}{2},0,0)$ &
$\bar{f}_2(\bar{5},-\frac{3}{2},0,-\frac{1}{2},0,0)$ &
$\ell_2^c(1,\frac{5}{2},0,-\frac{1}{2},0,0)$ \\
 
$F_3(10,\frac{1}{2},0,0,\frac{1}{2},-\frac{1}{2})$ &
$\bar{f}_3(\bar{5},-\frac{3}{2},0,0,\frac{1}{2},\frac{1}{2})$ &
$\ell_3^c(1,\frac{5}{2},0,0,\frac{1}{2},\frac{1}{2})$ \\
 
$F_4(10,\frac{1}{2},-\frac{1}{2},0,0,0)$ &
$f_4(5,\frac{3}{2},\frac{1}{2},0,0,0)$ &
$\bar\ell_4^c(1,-\frac{5}{2},\frac{1}{2},0,0,0)$ \\
 
$\bar{F}_5(\overline{10},-\frac{1}{2},0,\frac{1}{2},0,0)$ &
$\bar{f}_5(\bar{5},-\frac{3}{2},0,-\frac{1}{2},0,0)$ &
$\ell_5^c(1,\frac{5}{2},0,-\frac{1}{2},0,0)$ \\
\hline
\end{tabular}
\label{table:4}
 
\vspace*{0.5 cm}

\begin{tabular}{|l||l||l|}
\hline
$h_1(5,-1,1,0,0,0)$ & $h_2(5,-1,0,1,0,0)$ & $h_3(5,-1,0,0,1,0)$ \\
$h_{45}(5,-1,-\frac{1}{2},-\frac{1}{2},0,0)$ & &  \\
\hline
\end{tabular}

\vspace*{0.5 cm}

\begin{tabular}{|l||l||l|}
\hline
$\phi_{45}(1,0,\frac{1}{2},\frac{1}{2},1,0) $ &
$\phi_{+}(1,0,\frac{1}{2},-\frac{1}{2},0,1) $ &
$\phi_{-}(1,0,\frac{1}{2},-\frac{1}{2},0,-1) $ \\
$\Phi_{23}(1,0,0,-1,1,0)$ &
$\Phi_{31}(1,0,1,0,-1,0)$  &
$\Phi_{12}(1,0,-1,1,0,0)$ \\
$\phi(1,0,\frac{1}{2},
-\frac{1}{2},0,0)$ &
$\Phi(1,0,0,0,0,0)$ & \\
\hline
\end{tabular}
 
\vspace*{0.5 cm}

\begin{tabular}{|l||l||l|}
\hline
$\Delta_1(0,1,6,0,-\frac{1}{2},\frac{1}{2},0)$ &
$\Delta_2(0,1,6,-\frac{1}{2},0,\frac{1}{2},0)$ &
$\Delta_3(0,1,6,-\frac{1}{2},-\frac{1}{2},0,
\frac{1}{2})$ \\
$\Delta_4(0,1,6,0,-\frac{1}{2},\frac{1}{2},0)$ &
$\Delta_5(0,1,6,\frac{1}{2},0,-\frac{1}{2},0)$ & \\
\hline 
\end{tabular}

\vspace*{0.5 cm}

\begin{tabular}{|l||l||l|}
\hline
$T_1(0,10,1,0,-\frac{1}{2},\frac{1}{2},0)$ &
$T_2(0,10,1,-\frac{1}{2},0,\frac{1}{2},0)$ &
$T_3(0,10,1,-\frac{1}{2},-\frac{1}{2},0,\frac{1}{2})$ \\
$T_4(0,10,1,0,\frac{1}{2},-\frac{1}{2},0)$ &
$T_5(0,10,1,-\frac{1}{2},0,\frac{1}{2},0)$ & \\
\hline
\end{tabular}

\vspace*{0.2 cm}
\end{center}

{\small Table 3: 
{\it The chiral superfields are listed with their
quantum numbers \cite{aehn}.
The $F_i$, $\bar{f}_i$, $\ell_i^c$,
as well as the
$h_i$, $h_{ij}$ fields and the singlets
are listed with their
$ SU(5) \times U(1)' \times U(1)^4$ 
quantum numbers. 
Conjugate fields have opposite $U(1)' \times U(1)^4$
quantum numbers.
The fields $\Delta_i$ and $T_i$ are tabulated in terms
of their $U(1)' \times SO(10) \times SO(6) \times U(1)^4$
quantum numbers. }
}
\end{table}
 
As can be seen, the matter and
Higgs fields in this string model carry additional charges under
additional $U(1)$ symmetries~\cite{aehn}. There exist various
singlet fields, and  hidden-sector 
matter fields which transform
non-trivially under the $SU(4)\times SO(10)$ gauge symmetry,
some as sextets under $SU(4)$, namely $\Delta_{1,2,3,4,5}$, and some as
decuplets under $SO(10)$, namely $T_{1,2,3,4,5}$. There are also
quadruplets of the $SU(4)$ hidden symmetry which possess fractional
charges. However,  these are confined and
will not concern us further.

The usual flavour assignments of the light
Standard Model particles in this model are as follows:
\bea
\bar{f}_1 : \bar{u}, \tau, \; \; \;
\bar{f}_2 : \bar{c}, e/ \mu, \; \; \;
\bar{f}_5 : \bar{t}, \mu / e \nonumber \\
F_2 : Q_2, \bar{s}, \; \; \;
F_3 : Q_1, \bar{d}, \; \; \;
F_4 : Q_3, \bar{b} \nonumber \\
\ell^c_1 : \bar{\tau}, \; \; \;
\ell^c_2 : \bar{e}, \; \; \;
\ell^c_5 : \bar{\mu}
\label{assignments}
\eea
up to mixing effects, which are discussed in more detail in
\cite{ELLN}.
We chose non-zero vacuum expectation values for the 
following singlet and hidden-sector fields:
\beq
\Phi_{31}, \bar{\Phi}_{31}, \Phi_{23}, 
\bar\Phi_{23},\phi_2, \bar{\phi}_{3,4},  \phi^-, 
 \bar\phi^+ ,  \phi_{45},  \bar{\phi}_{45}, \Delta_{2,3,5}, T_{2,4,5}
\label{nzv}
\eeq
The vacuum expectation values of the hidden-sector fields
must satisfy the additional constraints 
\beq
 T_{3,4,5}^2=T_i\cdot T_4=0,\,\, \Delta_{3,5}^2=0,\,\, T_2^2+\Delta_2^2=0
\label{hcon}
\eeq
For further discussion, see~\cite{ELLN} and references therein.


\end{document}